\documentclass[11pt]{article}
\usepackage{moriond,epsfig}

\def\PRD{{\em Phys. Rev.} D}

\def\be{\begin{equation}}
\def\ee{\end{equation}}
\def\bea{\begin{eqnarray}}
\def\eea{\end{eqnarray}}

\begin{document}
\vspace*{4cm}  
\title{FIRST DATA FROM D\O~IN RUN 2}

\renewcommand{\thefootnote}{\fnsymbol{footnote}}
\author{M. VERZOCCHI \footnote{Representing the D\O~Collaboration}}

\address{Department of Physics, University of Maryland,\\
College Park, MD 20742, USA}

\maketitle\abstracts{
Run 2 of the Tevatron collider at Fermilab has begun in the spring of 2001. 
During its first year of operation the Tevatron has delivered an integrated
luminosity of approximately 30 $\mathrm{pb}^{\mathrm{-1}}$\ to the D\O~experiment. 
These data have been used to commission the detector. The progress in understanding 
the detector performance and the prospects for an exciting physics program to 
be carried out in the next years are the subject of this review.}

\renewcommand{\thefootnote}{\alph{footnote}}

\section{Physics goals for Run 2 at the Tevatron}
\label{sec:goals}

The Tevatron collider at Fermilab is the world's highest energy accelerator,
colliding protons and antiprotons at a centre of mass energy of $\sim\mathrm{2}$
TeV. Since the end of Run 1 the accelerator complex has been upgraded to
raise the collision energy and to deliver larger integrated luminosities
(2 $\mathrm{fb}^{\mathrm{-1}}$~before 2005, 
and 15 $\mathrm{fb}^{\mathrm{-1}}$~before the startup
of the LHC), extending the physics reach of the experiments. 
The D\O~detector has also been upgraded~\cite{upgrade} to cope with the reduced time 
between bunch crossings, the increase in luminosity and backgrounds, and
to extend the physics capabilities of the experiment.

The physics goals of Run 2 include the investigation of the electroweak 
symmetry breaking mechanism and searches for physics beyond the Standard Model. 
The luminosity 
gain will  also yield large statistics for precise measurements at lower mass 
scales, such as {\it b}--physics and QCD. 

The large increase in luminosity means that the reach for discoveries at
the highest mass scales will be increased, while formerly rare processes, like the 
production of weak bosons and of the top quark, become the object
of precision measurements. With 15 $\mathrm{fb}^{\mathrm{-1}}$~at the end of Run 2, 
the Tevatron experiments will be able to search for the Higgs boson in much 
of the phase space currently allowed by Standard Model fits~\cite{run2hig,lepewwg}.
Meanwhile this allowed range can be reduced with improved precise measurements of 
the masses of the top quark and of the W boson, reducing the uncertainty in their
determination respectively to 1--2 GeV and 20 MeV at the end of Run 2~\cite{ewrun2}.

With 2 $\mathrm{fb}^{\mathrm{-1}}$ the mass reach in the search for supersymmetric 
particles will extend to 400 GeV for  squark and gluinos, 200 GeV for stop 
and sbottom quarks and 180 GeV for charginos in the trilepton final state~\cite{susy}.
Other models (TeV--scale gravity and extra dimensions, technicolour, leptoquarks) will also
be used to guide searches for new phenomena. In addition to these optimised searches,
model independent searches for new physics will also be performed, following the
approach pioneered by D\O~in Run 1~\cite{sleuth}.

\section{Upgrade of the D\O~detector}
\label{sec:upgrade}

\begin{figure}[h]
\begin{center}
\epsfig{figure=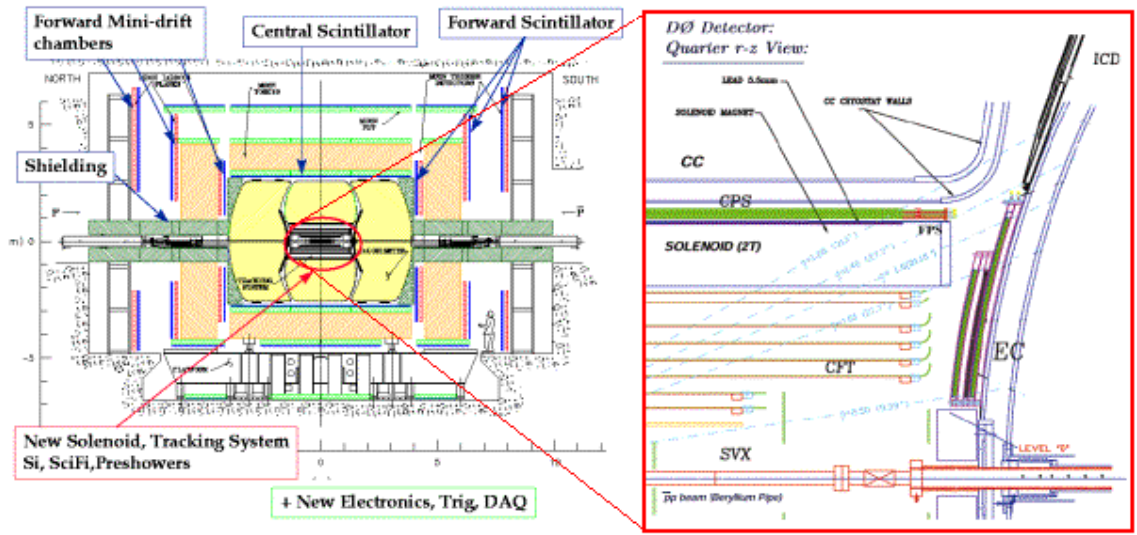,width=0.95\textwidth}
\caption{Transverse view of the D\O~detector, highlighting the parts
of the detector which have been upgraded in view of Run 2.}
\label{fig:d0dete}
\end{center}
\end{figure}

The Run 2 upgrade builds on the strengths of the Run 1 D\O~detector, its
state of the art hermetic calorimeter system and its lepton
identification capabilities over a large rapidity range. To achieve the
Run 2 physics goals the detector has undergone a series of large changes,
highlighted in the cross sectional view of the D\O~detector shown in
Fig.~\ref{fig:d0dete}.

A 2 T superconducting solenoid has been installed in front of the calorimeter
cryostat, surrounding a new tracking system comprising silicon microstrip
and scintillating fibres detectors. Preshower detectors have been installed
between the solenoid and the calorimeter in the central part of the detector,
and in front of the forward calorimeter to compensate for the energy loss
of electrons and photons in the solenoid and to improve the angular resolution
for photons. The forward muon system has been completely rebuilt, separating
the triggering function (using scintillator tiles), from the precision
tracking (using mini-drift tubes). The forward muon detector benefits from
the new shielding of the beam line, resulting in a large reduction of
backgrounds compared to Run 1. In the central region of the detector 
additional layers of scintillators have been added to the muon system,
allowing a more extensive  reduction of out of time backgrounds.
The readout electronics, the trigger and DAQ systems have all been rebuilt
to cope with the reduced time between crossings and the increase in luminosity.

\newpage
\section{Detector performance}
\label{sec:perform}

\subsection{Calorimeter}
\label{sub:calo}

\begin{figure}[ht]
\begin{center}
\begin{tabular}{lr}
\epsfig{figure=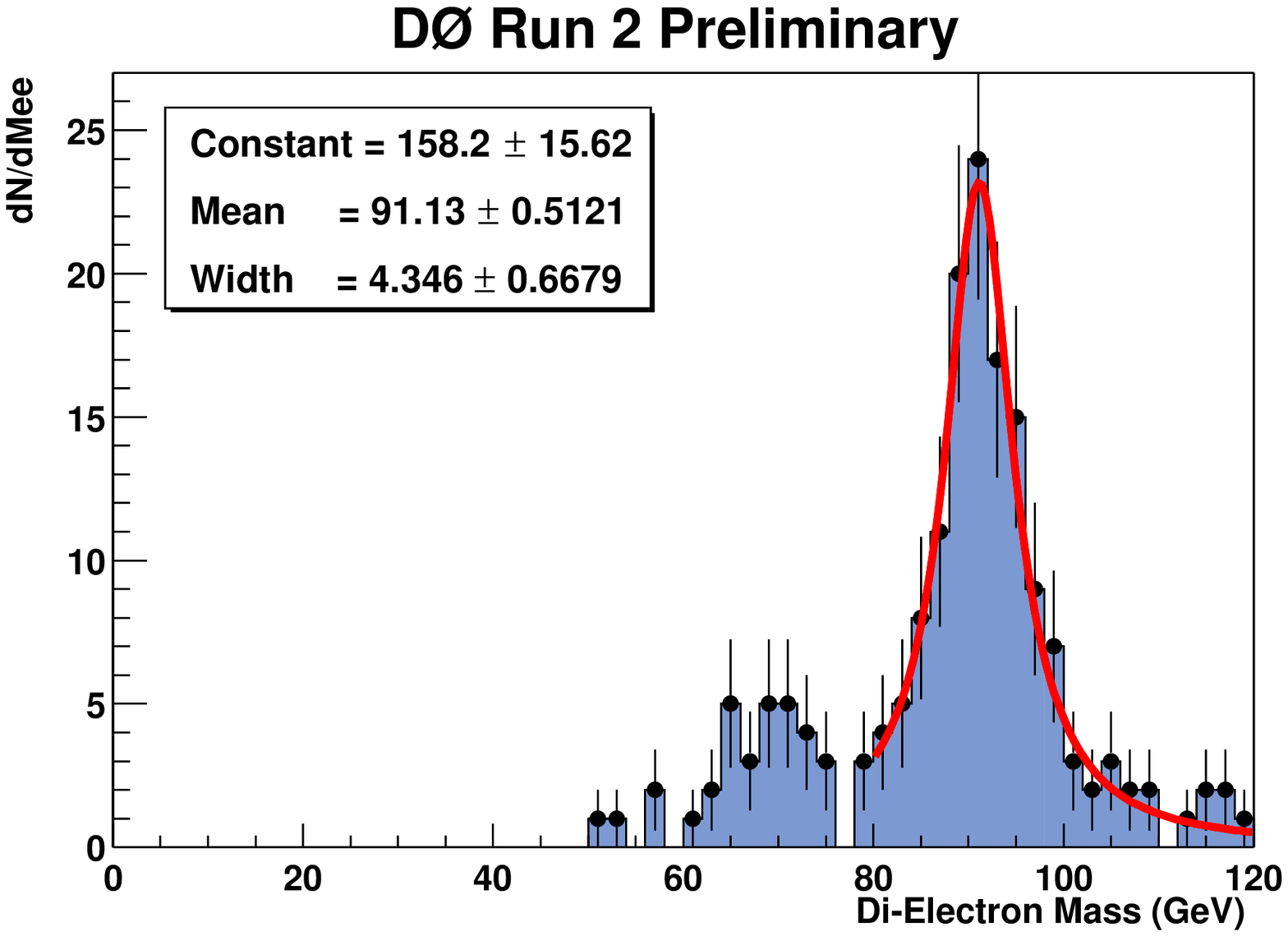,width=0.56\textwidth} &
\epsfig{figure=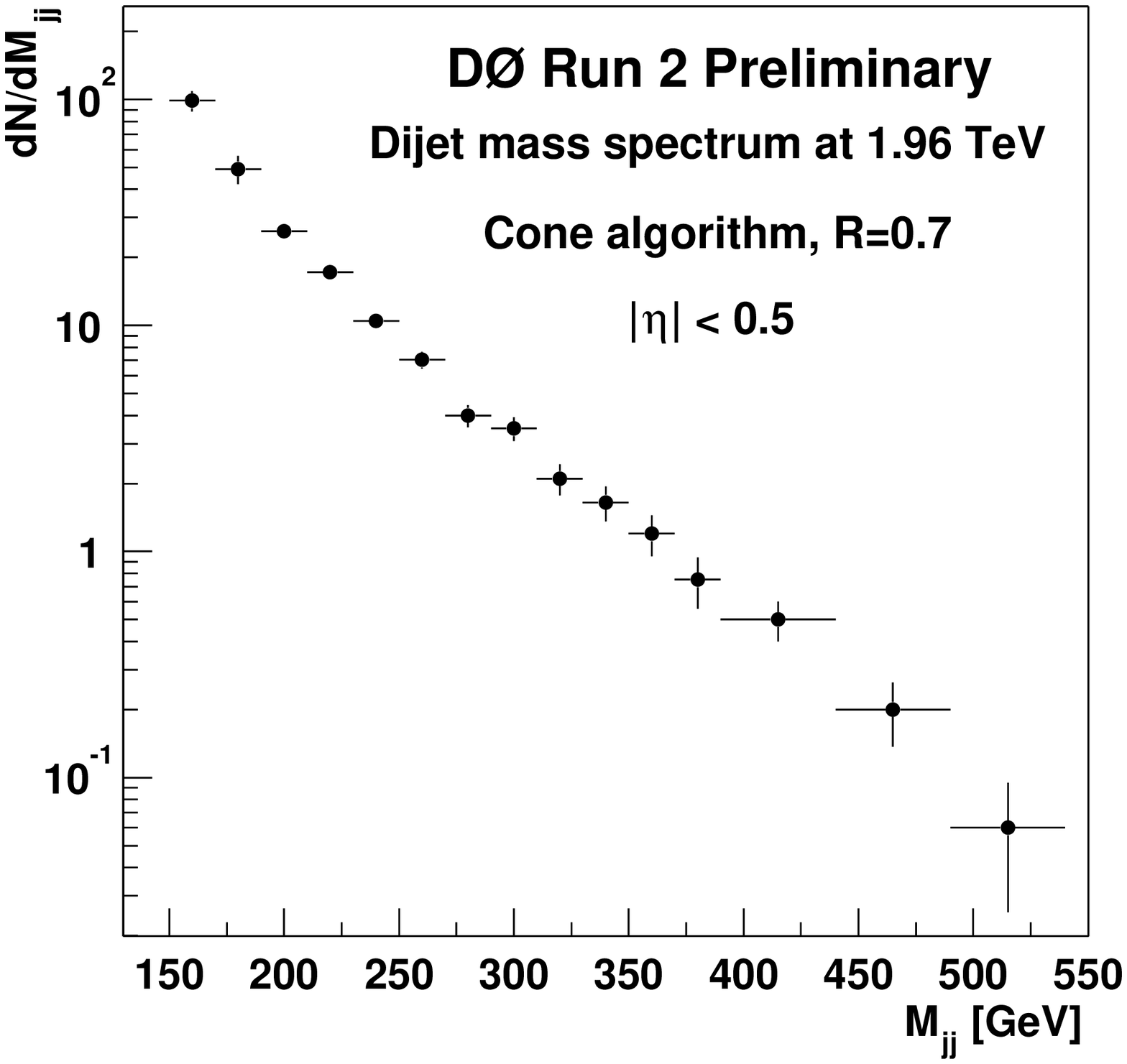,width=0.39\textwidth} 
\end{tabular}
\caption{Invariant mass distribution for events with two high $\mathit{p}_{\mathrm{T}}$ 
electrons (left) and for two jet events (right).}
\label{fig:calo}
\end{center}
\end{figure}

The D\O~calorimeter is a 55k channels U/LAr calorimeter with fine longitudinal
and transverse segmentation,
uniform response and good energy resolution. New readout electronics with
analog pipelines has been installed for Run 2. The calorimeter has been 
fully operational since the beginning of the run with less than 0.1\% bad channels.  
The commissioning of the preshower detectors and of the new inter--cryostats 
detectors (scintillator tiles installed in the gaps between the barrel and 
endcap calorimeters to improve the resolution on the missing transverse momentum) 
was still ongoing at the time of the conference. A preliminary calibration of
the calorimeter has been obtained investigating the invariant mass spectrum of 
dielectron events originating from the decay of $\mathrm{Z}^{\mathrm{0}}$ bosons,
shown in Fig.~\ref{fig:calo}. The knowledge of the absolute calorimeter
calibration is currently limited by the available statistics in the
$\mathrm{Z}^{\mathrm{0}}$ sample.  

  The energy scale for jets was then determined using $\gamma\mathrm{+jets}$ 
events. Also shown in Fig.~\ref{fig:calo} is a preliminary dijet invariant 
mass spectrum, which does not include correction for trigger threshold
effects. Due to the increased centre of mass energy of the Tevatron
collider, D\O~has already observed events at invariant masses in excess
of 500 GeV in a sample corresponding to an integrated luminosity
of only about 1 $\mathrm{pb}^{\mathrm{-1}}$.

  While the resolution on the missing transverse momentum will improve with
a better understanding of the calibration of the calorimeter,
the Monte Carlo already provides a good description of the significance
of the measurement of the missing transverse momentum, provided the
error on this quantity is taken from data. This permits the
use of the missing transverse momentum for analyses, one of the most
important tools in the searches for physics beyond the Standard Model.

\subsection{Muon detector}
\label{sub:muon}

In D\O~the trajectories of muons penetrating the iron toroid 
surrounding the calorimeters are measured by drift tubes. Trigger 
signals are obtained from scintillator counters, which provide a timing
signal used to reject background from cosmic rays. In the central region the
Run 1 drift tubes are used, with a faster gas mixture and
new readout electronics, and new scintillator counters have been
installed. The forward system has been completely rebuilt for Run 2:
it includes 3 layers of mini drift tubes and scintillator pixel
counters. The muon system has been fully operational since the beginning of
Run 2 and thanks to the new shielding close to the beam line 
the detectors can easily discriminate muons from the low rate of
out of time backgrounds.

\begin{figure}[ht]
\begin{center}
\begin{tabular}{lr}
\epsfig{figure=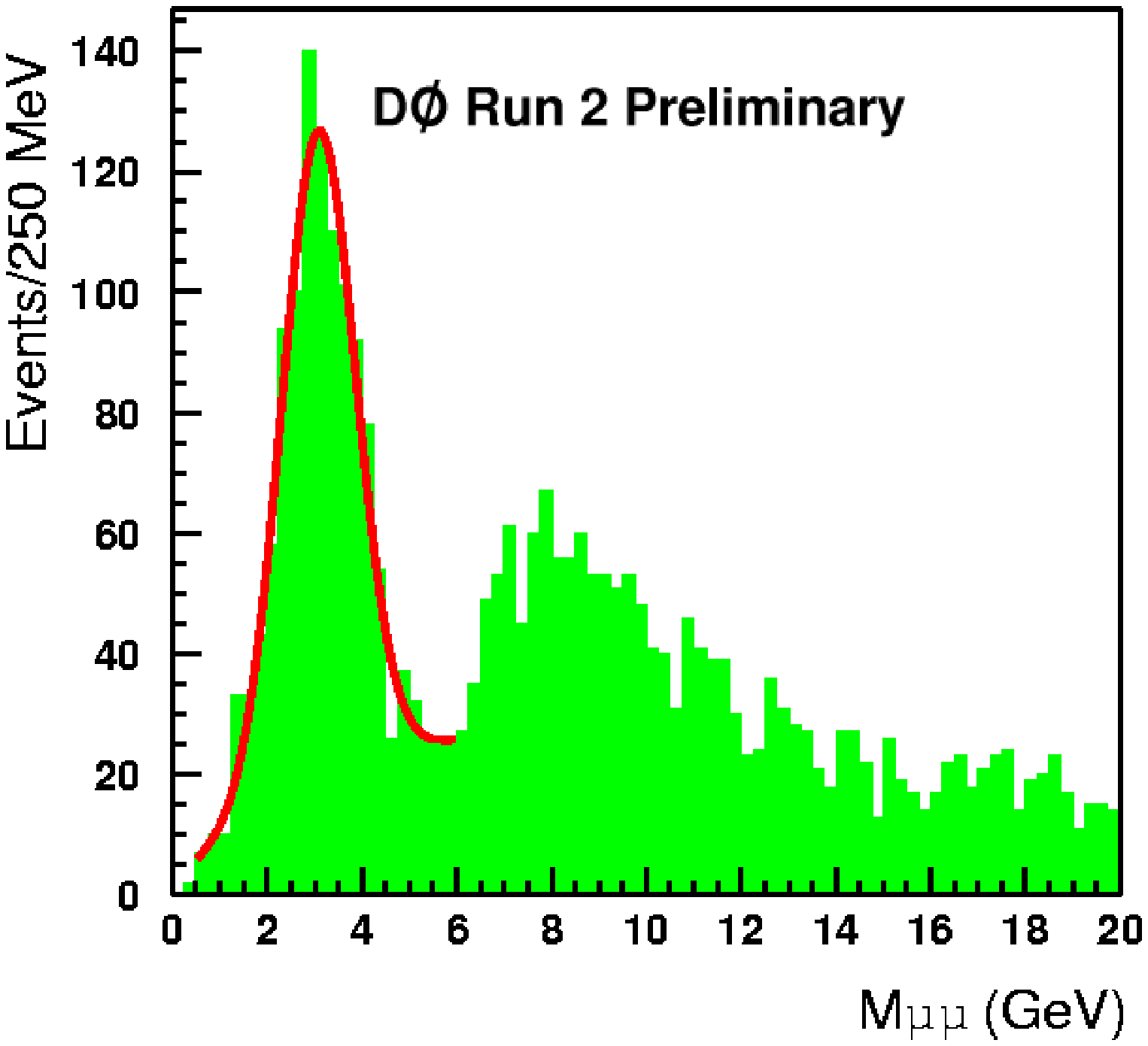,width=0.40\textwidth} &
\epsfig{figure=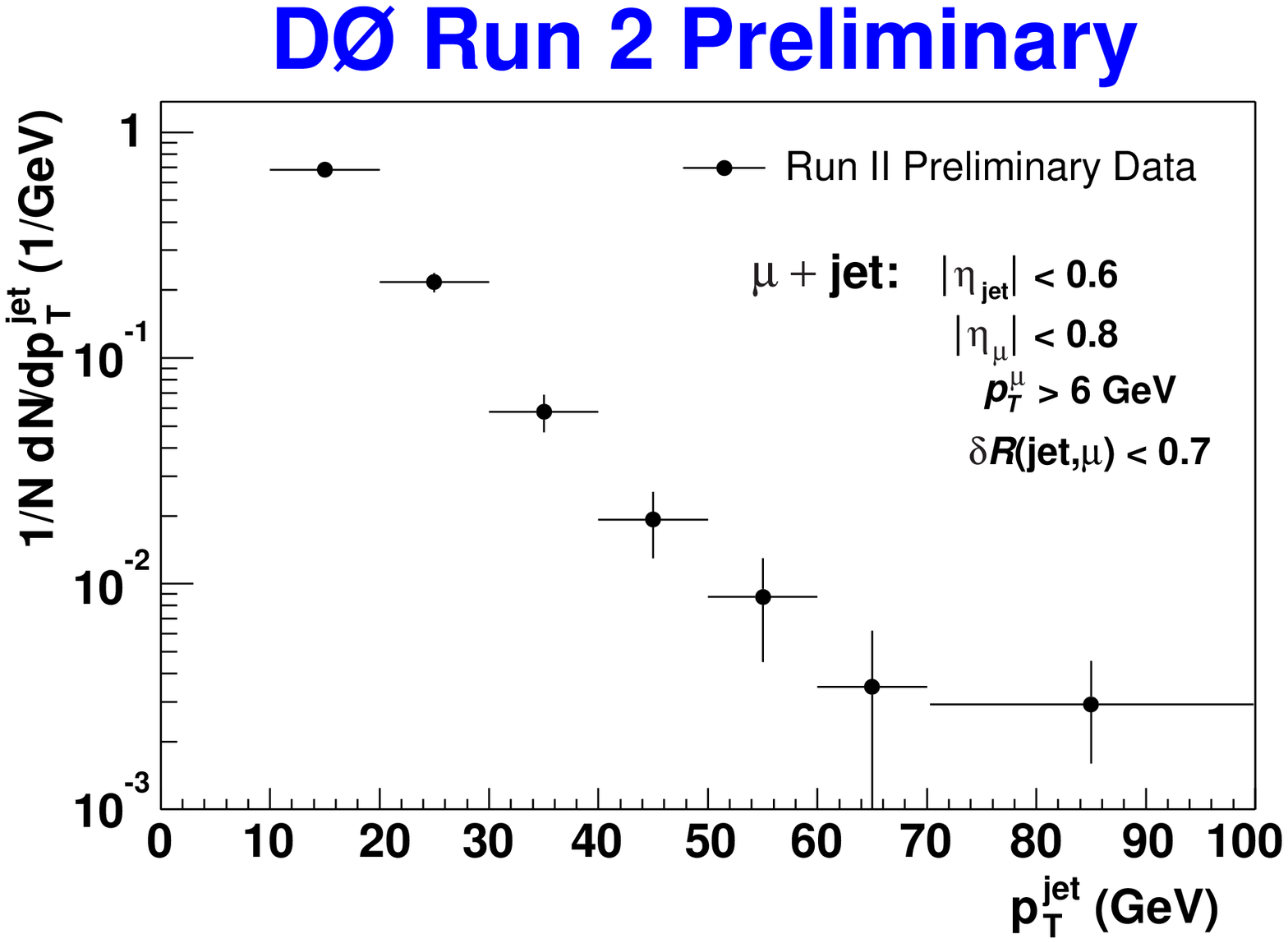,width=0.55\textwidth} 
\end{tabular}
\caption{Invariant mass distribution for dimuon events, using
only the muon momentum measurement in the iron toroid (left).
Transverse momentum spectrum for jets with a soft muon
{\it b}--tag (right).}
\label{fig:muon}
\end{center}
\end{figure}

Two examples of the measurements performed so far to study the
performance of the muon detector are shown in Fig.~\ref{fig:muon}.
The left plot shows the invariant mass distribution for muons
having $\mathit{p}_{\mathrm{T}}>\mathrm{3}$ GeV. The
signal from the $\mathrm{J}/\psi$ resonance is clearly visible
above the background. Its width is consistent with expectations
for the muon system only, where the momentum resolution is dominated by the multiple
scattering in the iron toroid. The muon momentum resolution improves
with the use of the information obtained from the
central tracking system (see Sec.~\ref{sub:track}). The right plot shows the spectrum
of jets having a soft--muon {\it b}--tag. Muons having a transverse 
momentum relative to the jet axis in excess of 1 GeV most 
likely come from the decay of a {\it b} quark. Both the transverse
momentum distribution of muons relative to the jet and the 
transverse momentum distribution of these jets agree with expectations,
based on the Run 1 data.

\subsection{Tracker}
\label{sub:track}

\begin{figure}[ht]
\begin{center}
\begin{tabular}{lr}
\epsfig{figure=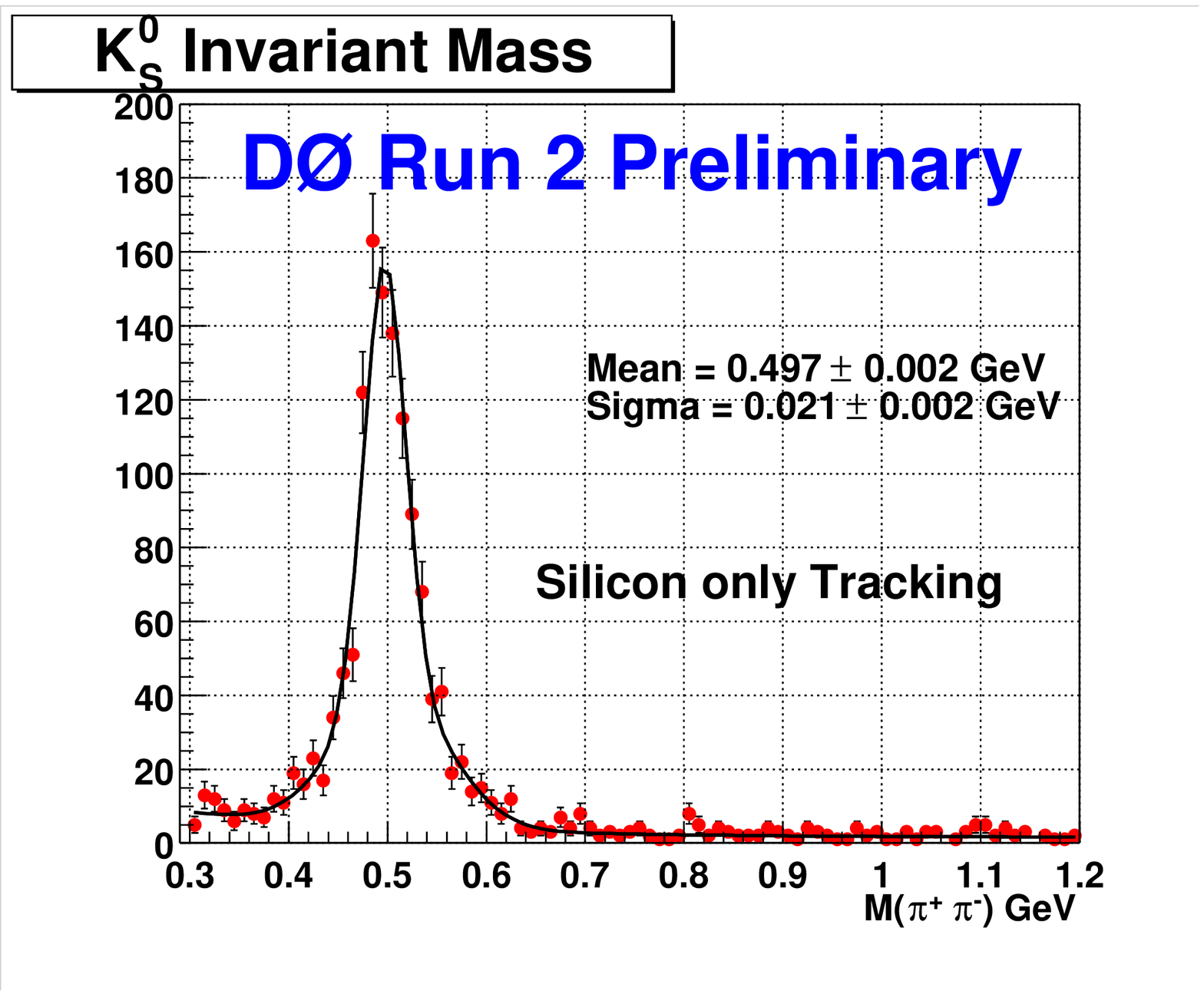,width=0.53\textwidth} &
\epsfig{figure=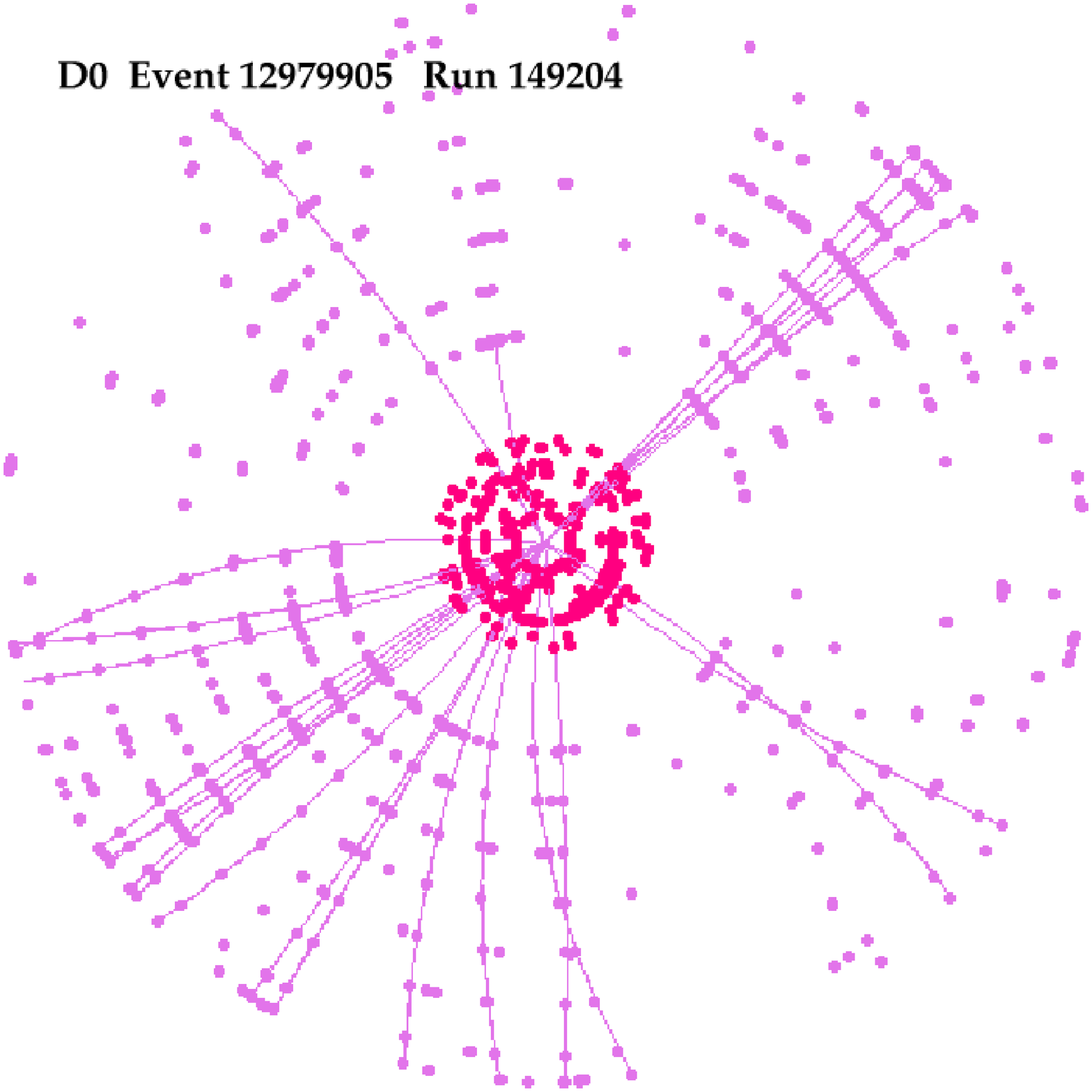,width=0.42\textwidth}
\end{tabular}
\caption{Invariant mass distribution for pairs of unlike sign tracks,
using tracks reconstructed only in the silicon detector (left). 
Cross sectional view of one event in the tracking system 
(scintillating fibres and silicon microstrip detector): only
tracks with $\mathit{p}_{\mathrm{T}}>\mathrm{500}$ MeV
are reconstructed (right).}
\label{fig:track}
\end{center}
\end{figure}

The D\O~tracking system is entirely new: it is based on a silicon microstrip tracker
(SMT) and a central fibre tracker (CFT) installed inside a 2 T solenoid. The 
SMT consists of 4 barrel layers of single and double--sided silicon microstrip
detectors, interspread with disks arranged perpendicular to the beam
direction. These, together with additional disks in the forward directions,
allow efficient track reconstruction in the SMT up to $|\eta|=\mathrm{2.5}$
independently from the position of the primary vertex, which has a gaussian
distribution along the beam axis with a RMS of 30 cm. Altogether the SMT comprises $\sim$800k
readout channels. It is has been in continuous operation since
the beginning of Run 2. Fig.~\ref{fig:track} shows the 
$\mathrm{K}^{\mathrm{0}}_{\mathrm{s}}$ peak obtained from
the invariant mass of unlike sign tracks reconstructed in the SMT system alone.

At larger radii (between 20 and 51 cm) tracking is performed in 8 double
layers of 840 $\mu$m diameter scintillating fibres. Each layer has
two axial and two $\mathrm{2}^o$ stereo fibres, read out through
visible light photon counters operating at 9 K, with 85\% quantum 
efficiency and good signal to noise ratio. The readout electronics
for the fibre tracker (and the preshower detectors) has been completely
installed in the spring of 2002. Fig.~\ref{fig:track} shows the hits and the 
reconstructed tracks in the D\O~tracker for a typical two jet event.
Track information is already being used in
analyses, improving the resolution of the muon measurement, and providing
a useful tool for the calibration of the electromagnetic calorimeter.
Work is underway to improve the understanding of the vertexing algorithms
and to develop impact parameter and vertex based {\it b} quark tags, which
are crucial in the search for the Higgs boson and for reducing the
background in top analyses.

\subsection{DAQ and trigger}
\label{sub:trigdaq}

The trigger and DAQ systems have been almost completely rebuilt for Run 2.
High $\mathrm{p}_{\mathrm{T}}$ triggers have been running without prescales
during the first year of Run 2 operations despite limitations in the D\O~data 
taking capabilities due to delays in the delivery of L2 
CPUs and of L3 components. Most of the L2 triggers and a new Ethernet based DAQ
system are being commissioned: D\O~will be capable of handling design trigger 
rates before the beginning of the summer. The L1 central tracking trigger (CTT),
which uses the track measurements in the CFT, is being commissioned. Its use will 
result in sharper turn--on curves for muon triggers. The L2 silicon track trigger, 
an important addition for Higgs physics, will be installed and commissioned later in the fall.
 
\section{First physics results}

\begin{figure}[ht]
\begin{center}
\begin{tabular}{lr}
\epsfig{figure=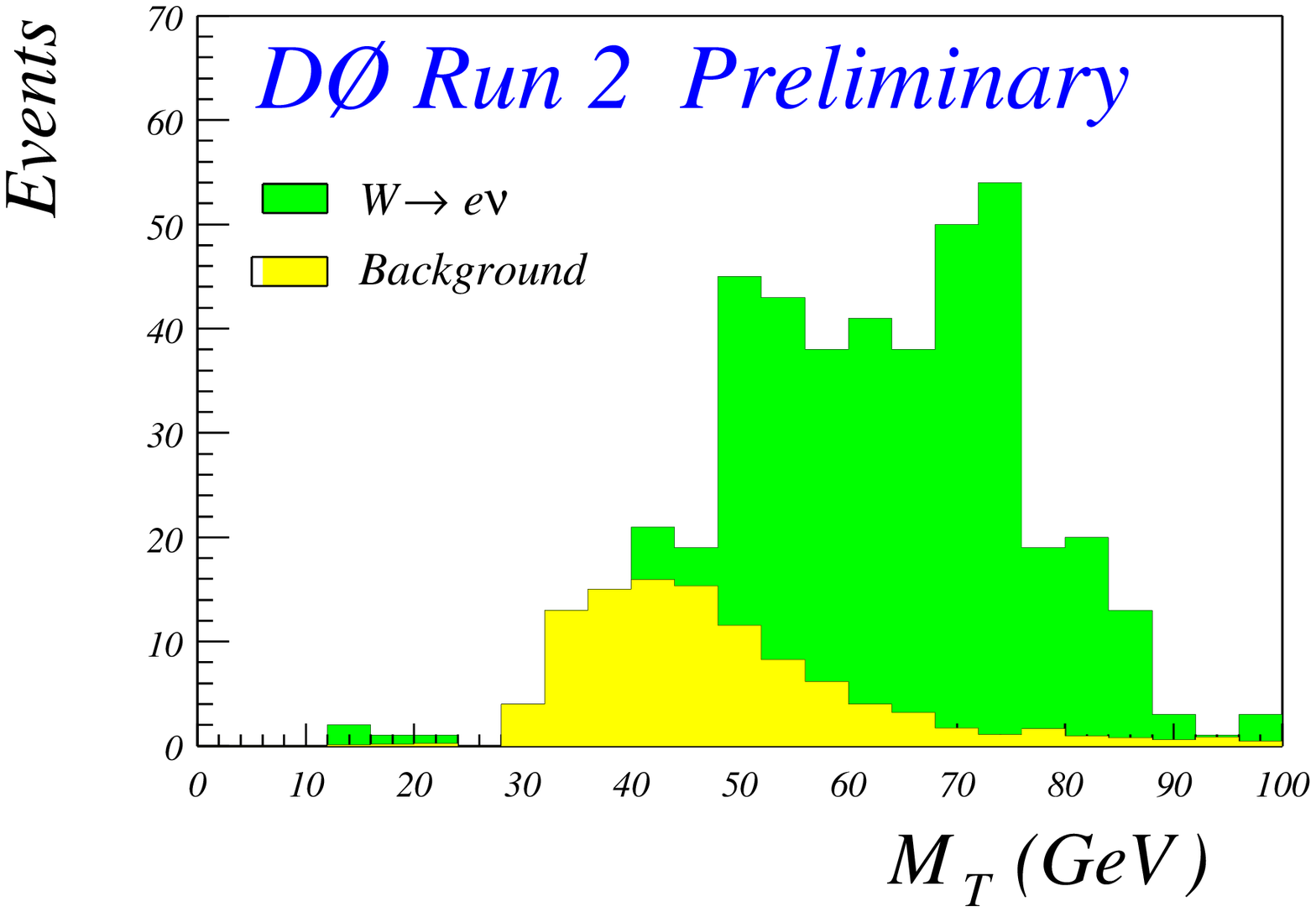,width=0.56\textwidth} &
\epsfig{figure=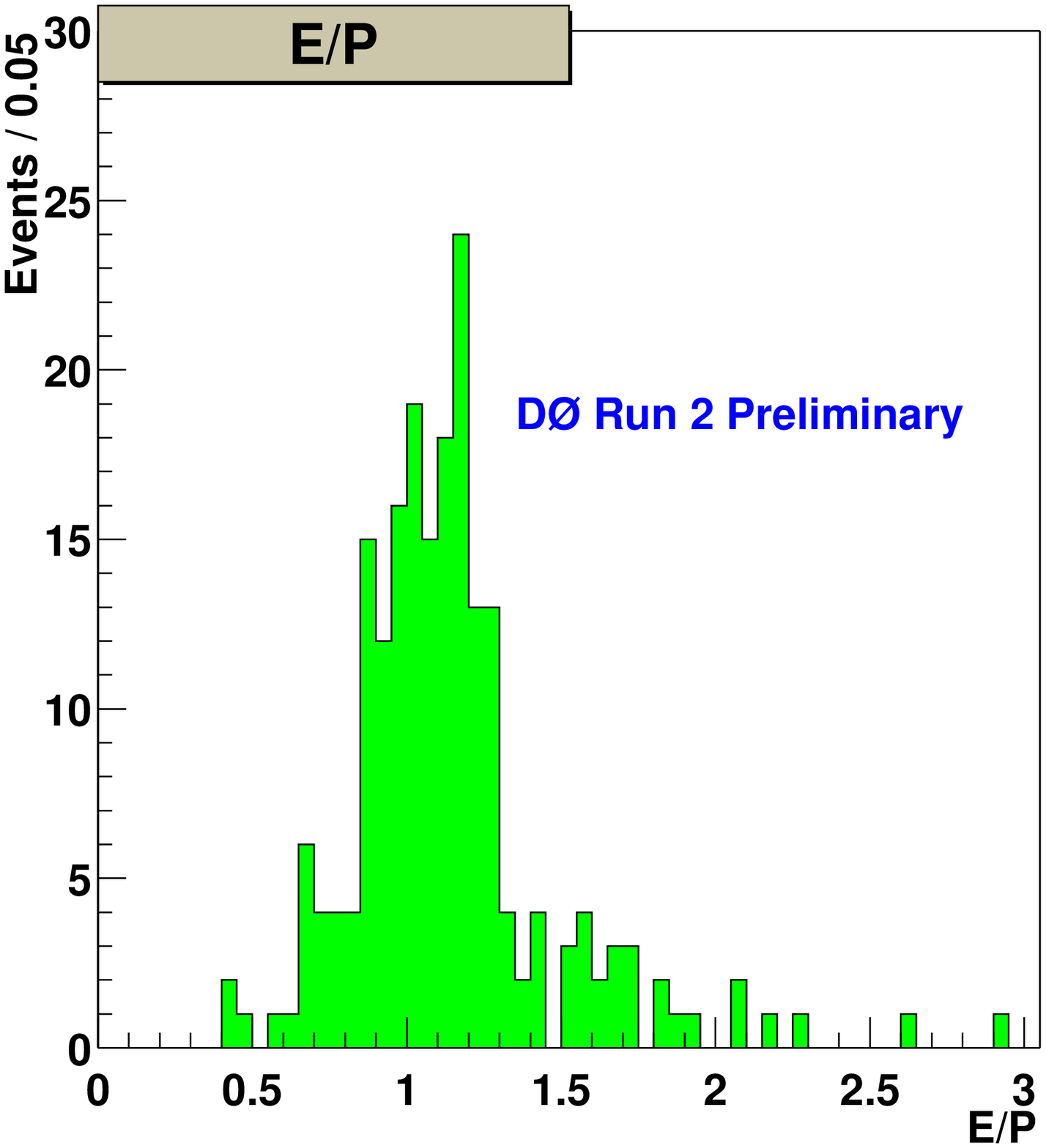,width=0.39\textwidth}
\end{tabular}
\caption{Transverse mass distribution for a sample of $\mathrm{W}\rightarrow\mathrm{e}\nu$
candidate events (left) and distribution of E/p for candidate electrons, using 
preliminary alignment and calibration constants (right).}
\label{fig:w}
\end{center}
\end{figure}

In addition to the physics signals already discussed in the previous
section, a relatively clean sample of $\mathrm{W}\rightarrow\mathrm{e}\nu$ 
candidates was obtained by selecting events with a high $\mathrm{p}_{\mathrm{T}}$ 
electromagnetic cluster matched to a track and large missing transverse energy
(see Fig.~\ref{fig:w}). The background has been estimated from data and consists 
mainly of QCD events with fake electrons. Also shown in Fig.~\ref{fig:w} is
the E/p ratio for the candidate electrons.  Those W candidates with additional
jets will constitute the main background for top and Higgs analyses. 

Other 
preliminary results include the first candidates (most likely from background
sources) in searches for trileptons and leptoquarks. More extensive results
using higher quality data and larger luminosities are expected for the
summer.  

\section{Conclusions}
\label{sec:concl}

The luminosity delivered by the Tevatron in the first year of Run 2
has been used by the D\O~experiment mainly for detector commissioning 
purposes, allowing enormous progress in the understanding of the detector
performance. Preliminary analyses have been performed using a subset of
the delivered luminosity, indicating that the D\O~collaboration will be able
to fully exploit the physics opportunities presented by Run 2.

\section*{Acknowledgements}
The author would like to thank all his D\O~colleagues, who have helped
in preparing the results, and the conference organisers for providing
such an enjoyable and stimulating atmosphere.

\end{document}